\documentclass[twocolumn,english,aps,10pt]{revtex4-1}
\usepackage[utf8x]{inputenc}
\usepackage{textcomp}
\usepackage{graphicx}
\usepackage{hyperref}
\usepackage{amssymb}
\usepackage{latexsym}
\usepackage{babel}

\usepackage{fancyhdr}
\fancypagestyle{firstpage}{%
\lhead{\footnotesize{This is the peer reviewed version of the following article: [Voronenkov, V., Bochkareva, N., Zubrilov, A., Lelikov, Y., Gorbunov, R., Latyshev, P. and Shreter, Y. (2019), Hydride Vapor-Phase Epitaxy Reactor for Bulk GaN Growth. Phys. Status Solidi A.], which has been published in final form at [\href{https://doi.org/10.1002/pssa.201900629}{doi:10.1002/pssa.201900629}]. This article may be used for non-commercial purposes in accordance with Wiley Terms and Conditions for Use of Self-Archived Versions.}}
  \rhead{}
}
\begin{document}

\title{Hydride Vapor-Phase Epitaxy Reactor for Bulk GaN Growth}

\author{Vladislav Voronenkov} \email{voronenkov@mail.ioffe.ru}
\author{Natalia Bochkareva}
\author{Andrey Zubrilov}
\author{Yuri~Lelikov}
\author{Ruslan Gorbunov}
\author{Philipp Latyshev}
\author{Yuri Shreter}

\affiliation{%
  Ioffe Institute, Politehnicheskaya~26, St.~Petersburg, 194021, Russia
}

\begin{abstract}
An HVPE reactor for the growth of bulk GaN crystals with a diameter of 50 mm was developed. Growth rate non-uniformity of 1\% was achieved using an axisymmetric vertical gas injector with stagnation point flow. Chemically-resistant refractory materials were used instead of quartz in the reactor hot zone. High-capacity external gallium precursor sources were developed for the non-stop growth of the bulk GaN layers. A load-lock vacuum chamber and a dry \textit{in-situ} growth chamber cleaning were implemented to improve the growth process reproducibility. Freestanding GaN crystals with a diameter of 50 mm were grown with the reactor.
\end{abstract}

\maketitle 
\thispagestyle{firstpage}

\section{Introduction}

The chloride-hydride epitaxy is the primary method of bulk GaN substrate production.
The HVPE method allows to grow epitaxial layers of high purity and to produce uncompensated semi-insulating material~\cite{fujikura2017sciocs}. The structural perfection of HVPE-grown layers is determined by the seed substrate quality: the dislocation density in the epitaxial layers grown on the bulk GaN substrate does not exceed the dislocation density in the substrate \cite{grzegory2006crystallization,bockowski2016challenges,imanishi2017-hvpe-on-na-flux}. The HVPE method is also promising for the epitaxy of high-purity device structures~\cite{fujikura2017sciocs}.

The high process temperature, together with the chemically aggressive substances involved in the HVPE process, makes the design of the reactor a challenging task.
Gallium, ammonia, and hydrogen chloride react to some extent with most heat-resistant materials, including quartz, often used in HVPE reactor design \cite{Tietjen01071966,maruska1969,safvi1997kuech,tavernier2000clarke,dam2004hageman-horizontal,segal2004illegems,monemar2005,richter2006aitron-horizontal,williams2007moustakas,fujito2009mchemical,amilusik2013topgan}, which leads to degradation of reactor parts and unintended doping of the grown crystal.
The exclusion of quartz from the hot zone of the reactor is known to reduce the level of unintentional doping significantly \cite{fujikura2017sciocs,fleischmann2019influence}.
The typical HVPE reactor layout with a gallium source located inside the hot zone of the reactor, near the growth chamber, imposes considerable restrictions on the growth chamber design, limits the source capacity, and complicates the reactor maintenance.

The process is usually performed at near-atmospheric pressures, which makes the gas flow prone to free convection and recirculation \cite{segal2004illegems,richter2006aitron-horizontal,hemmingsson2008modeling} and makes it difficult to obtain a uniform thickness distribution and supress the parasitic growth on the injector nozzle at the same time.

In this paper, a reactor designed for growing bulk GaN layers with a diameter of 2 inches is described.
To ensure long-term non-stop growth processes, a high-capacity external precursor source was developed, and the gas flow pattern in the growth chamber was optimized to suppress the parasitic growth at the injector nozzle completely. Heated exhaust lines and a large-capacity powder trap were used to prevent clogging. To reduce the time between growth processes, a vacuum load-lock chamber was used. 
 Chemically resistant materials were used in the hot zone of the reactor to reduce possible crystal contamination and to increase the lifetime of the reactor parts.
The uniformity of the deposition rate and the V/III ratio of less than 1\% was obtained.

\section{Growth system design}
\subsection{Modeling procedure details}
\begin{figure*}[htb]
	\includegraphics[width=\textwidth]{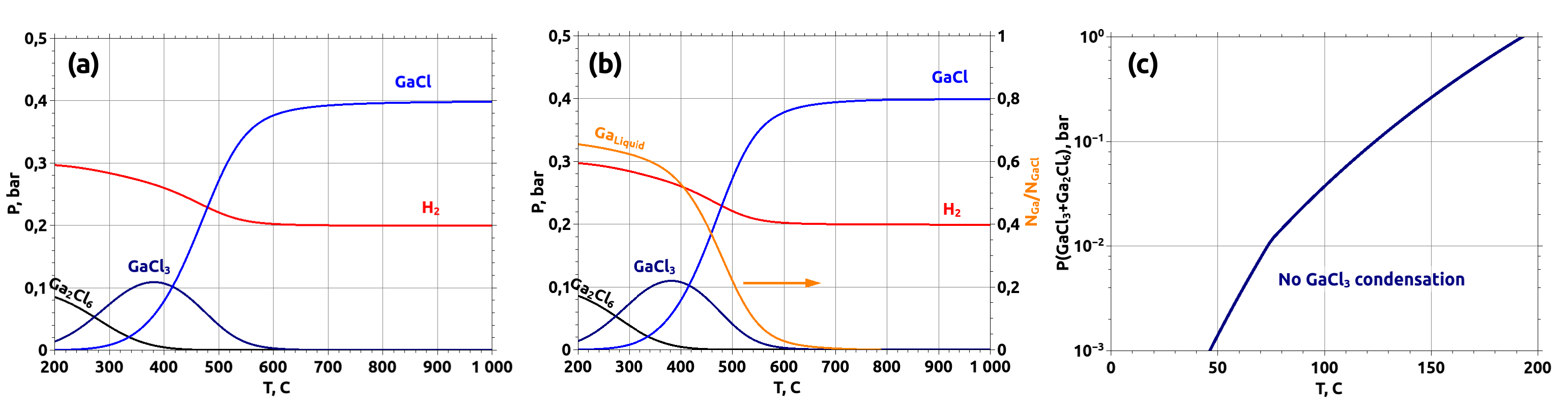}

	\caption{(a) The main products of gallium chlorination in a mixture of hydrogen chloride with inert gas at a ratio of 1:1. (b) Decomposition of gallium monochloride produced in a gallium boat at a temperature of 800~$^\circ$ C with a decrease in temperature. Liquid gallium is formed at temperatures lower than the chlorination temperature. (c) Equilibrium vapor pressure over condensed gallium trichloride.}
	\label{fig:boat}
\end{figure*}
\subsubsection{Chemical equilibrium analysis}
To calculate the chemical equilibrium, a hierarchical method was used \cite{villars1959method,colonna2004hierarchical,colonna2007improvements,vvv-thesis}.
Thermochemical properties of substances were sourced from \cite{gurvich1990thermodynamic,nist-janaf1998,przhevalskii1998thermodynamic,zinkevich2004thermodynamic,schafer1980mocl345,schonherr1988ptcl-growth,brewer1950chemistry}.
No preliminary assumptions were made about the composition of the system; 
all species for which thermochemical data were available in the sources \cite{gurvich1990thermodynamic,nist-janaf1998,przhevalskii1998thermodynamic,zinkevich2004thermodynamic,schafer1980mocl345,schonherr1988ptcl-growth,brewer1950chemistry} were accounted for in the calculation. Ideal gas behavior was assumed.
\subsubsection{CFD calculation}
The self-consistent calculation of gas flow, heat transfer, particle diffusion, and surface reactions was performed in a two-dimensional radially symmetric approximation by the finite volume method \cite{popinet2003gerris}. Surface reactions were taken into account under the assumption of a kinetically limited thermodynamic equilibrium: surface reactions were assumed to be infinitely fast except for the decomposition of ammonia into hydrogen and nitrogen, which was assumed to be kinetically inhibited. This approximation is applicable at typical GaN growth temperatures, which are in the range of 900--1100~$^\circ$C. The GaN deposition process at these temperatures is limited by diffusion through the boundary layer or thermodynamically limited \cite{seifert1981study,malinovsky1982growth,usui1997bulk,fornari2001hydride,koukitu2002surface}, and the process of ammonia decomposition on the surfaces is kinetically limited \cite{ban1972mass_gan,liu1978growth}.

\subsection{External boat}\label{ssec:external_boat}

The aim of implementing the external boat was to simplify the reactor design and service. External boat volume is not constrained by the dimensions of the reactor hot zone, and precursor refilling can be conducted without disassembling the reactor.  The main technical limitation when designing an external source is the need to transport gallium precursors at temperatures compatible with metal pipelines and conventional vacuum elastomeric or metal seals, i.e., at $T < 200$~$^{\circ}$C. This requirement can not be fulfilled using conventional chlorination process performed at temperatures above 600~$^{\circ}$C, where the main product of the chlorination reaction (Fig. \ref{fig:boat}a) is gallium monochloride,  formed by the reaction 
\begin{equation}
\rm Ga_{liquid} + HCl \rightarrow GaCl + \frac{1}{2}H_2,
\end{equation}
that decomposes with condensation of metallic gallium, at temperatures below the chlorination temperature (Fig. \ref{fig:boat}b).

One known way to overcome this problem is to use an organometallic  \cite{miura1995new,kryliouk1999large,fahle2013hcl} or organometallic chloride \cite{Lindeke1970-TEG-Cl,nickl1974preparation,ZHANG2000-TEG-Cl-GaN} precursor that is converted to gallium monochloride in the hot zone of the reactor by reaction with hydrogen chloride or by decomposition, respectively. The drawback of this approach is the possible carbon contamination and the formation of condensed carbon co-products \cite{park2005identification}.

Another possible way is to use gallium trichloride \cite{nickl1974preparation,lee1996vapor,varadarajan2004chloride,yamane2011tri},  that can be evaporated from a bubbler \cite{nickl1974preparation,lee1996vapor,varadarajan2004chloride}, or produced in a high-temperature boat by chlorination in excess of chlorine \cite{yamane2011tri}, or by chlorination at temperatures below 300~$^{\circ}$C, where the main products of the chlorination reaction are GaCl$_3$ and its dimer Ga$_2$Cl$_6$ (Fig. \ref{fig:boat}a) formed in the reaction:
\begin{equation}
\rm Ga_{liquid} + 3HCl \rightarrow GaCl_3 + \frac{3}{2}H_2
\label{ga-gacl3}
\end{equation}
Gallium trichloride condenses only at temperatures below 200~$^{\circ}$C (Fig. \ref{fig:boat}c), which makes it possible to transport it through pipes heated to moderate temperatures in the range of 100--200~$^\circ$C, depending on the degree of dilution by the carrier gas and the process pressure. 

Gallium trichloride can be used directly for the synthesis of GaN by the reaction 
\begin{equation}
\label{eq:GaCl3-deposition}
\rm GaCl_3 + NH_3 = GaN + 3HCl,
\end{equation}
however, high supersaturation of this reaction can lead to a parasitic reaction in the gas phase \cite{ueda2011effects}, also, the growth of the Ga-polar GaN surface is unstable due to the steric effect~\cite{murakami2016tri}.

To avoid this,  GaCl$_3$ can be reduced to GaCl by reaction with hydrogen in the hot zone \cite{kobayashi1991low}: 
\begin{equation}
\rm GaCl_3 + H_2 \rightarrow GaCl + 2HCl
\label{eq:GaCl3-decomposition}
\end{equation}
In this case, GaN deposition proceeds by the same reaction as in the reactor in the internal boat:
\begin{equation}
\rm GaCl + NH_3 \rightarrow GaN + HCl + H_2.
\label{eq:GAN-synthesis}
\end{equation}

Two types of external sources of gallium trichloride were successfully tested: an external boat with gallium chlorination at a temperature of about 200~$^\circ$C, and a bubbler evaporator filled with dry GaCl$_3$. In both cases, $\rm GaCl_3$ was delivered to the growth chamber through pipes heated to a temperature below 200~$^\circ$C. 

The external boat was a vacuum-tight ceramic vessel with metal flanges with an elastomeric seal.
Inside was a container filled with gallium; the surface area of liquid gallium was about 100~cm$^2$.
 The gallium container was heated to a temperature of about 200~$^\circ$C by an external resistive heater. During growth, 100 -- 300 sccm of HCl diluted in nitrogen was fed into the boat, which corresponded to a 33 -- 100 sccm GaCl$_3$ flow at the outlet of the external boat. The chlorination efficiency at a given temperature and flow was close to 100$\%$. The total flow at the exit of the boat was about 500~sccm. The entire system was thermally insulated to maintain the temperature of the walls and piping above the condensation point.

An alternative $\rm GaCl_3$ delivery method  using a bubbler evaporator has also been tested. The bubbler with a capacity of 1 liter was filled with gallium trichloride with a purity of 99.9999$\%$. Before the growth process, the bubbler was heated to a temperature above the melting point of $\rm GaCl_3$, usually 90 -- 110~$^\circ$C, which was maintained by a thermostat. The carrier gas was hydrogen or a mixture of nitrogen and hydrogen. The total gas flow at the bubbler outlet was also maintained at around 500~sccm. Like the external boat, the bubbler-based system was placed in a heated enclosure to prevent condensation. Although the bubbler system allowed to achieve results similar to that obtained using the external boat, it had much more stringent requirements for temperature stabilization accuracy and was more difficult to service.

\subsection{Growth chamber design}
\begin{figure*}[htb]
	\includegraphics[width=\textwidth]{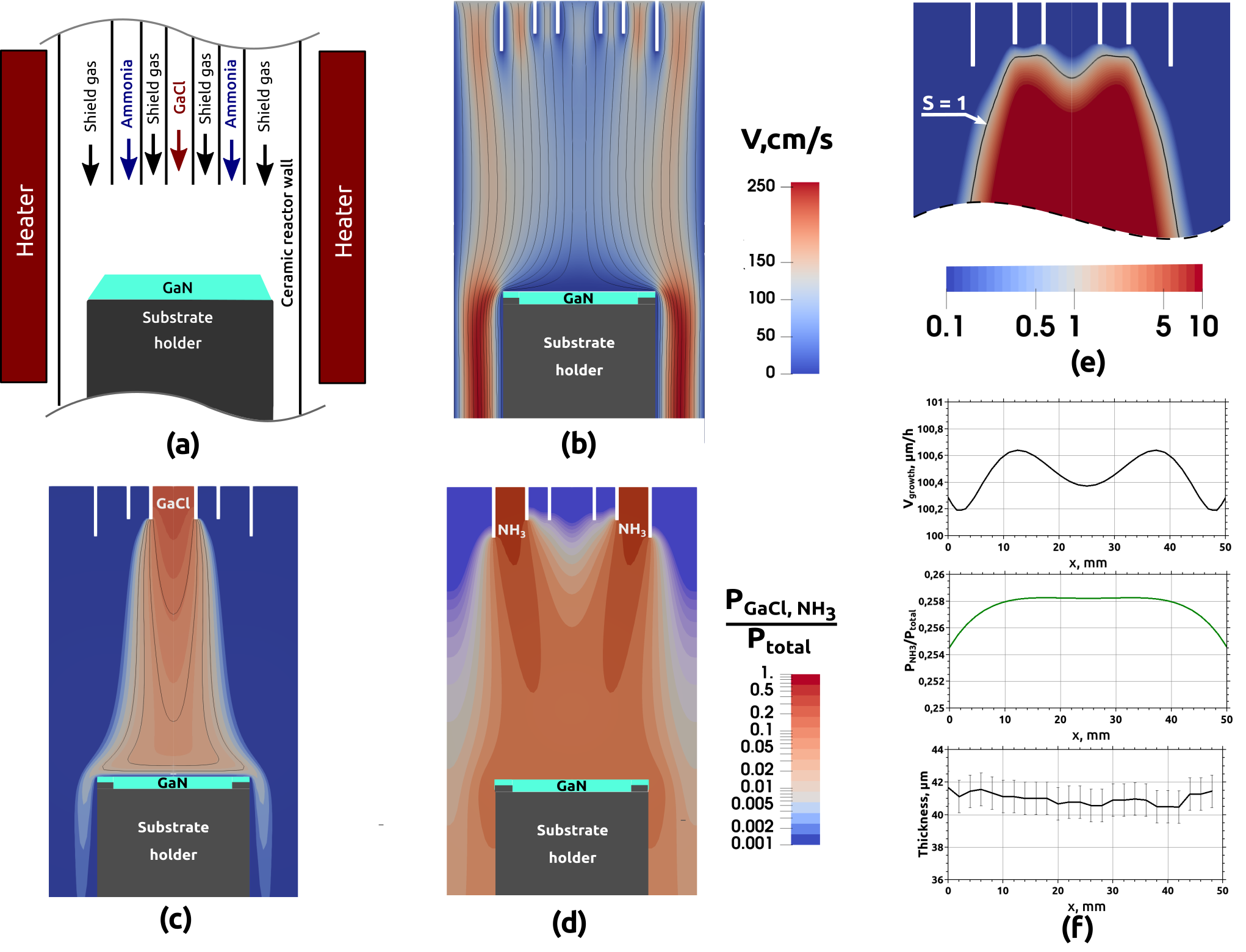}
	\caption{(a) Schematic	drawing of the reactor growth chamber. (b) Calculated velocity field. Color represents the magnitute of absolute gas velocity, streamlines show the laminar flow pattern without recirculation. (c) concentration field of GaCl. Lines of equal GaCl concentration near the substrate are near parallel to the surface, demonstrating a uniform boundary layer. (d) Concentration field of NH$_3$. (e) Relative saturation field near the injector nozzle. Black countour line represents the level of relative saturation $S = 1$. On the surface of the injector nozzle, $S < 1$, and parasitic GaN growth is thermodynamically unfavorable. (f) Distribution of the growth rate and ammonia concentration across the wafer surface, and a cross-section thickness profile of a 41~$\mu$m thick GaN film, measured by optical microscope.}
	\label{cfd}
\end{figure*}
The reaction chamber (Fig. \ref{cfd}a) is vertical with hot walls and external resistive heating that provides growth temperature up to 1500~$^\circ$C. All components of the growth chamber are made of non--oxide ceramics and refractory metals. The substrate with a diameter of 50 mm is placed horizontally, growth surface up. The gas injector consists of a set of coaxial tubes, implementing the stagnation point flow pattern (Fig. \ref{cfd}b). Nitrogen is used as a carrier gas. Gallium chloride is fed through the inner tube.

To completely prevent parasitic growth on the nozzle of the injector, a relative saturation $S$ of less than unity is maintained on the surface:
\begin{equation}
S = \frac{P_{GaCl}P_{NH_3}}{P_{HCl}{P_{H_2}}}K(T) < 1
\label{eq:supersaturation}
\end{equation}
where $P_x$ are the partial pressures of the corresponding substances, and $K(T)$ is the equilibrium constant of the GaN deposition reaction (\ref{eq:GAN-synthesis}). The inert gas shielding between the ammonia supply line and the gallium supply separates GaCl$_3$ and NH$_3$ and reduces $P_{GaCl}P_{NH_3}$ in the vicinity of the injector. To fulfill condition \ref{eq:supersaturation}, the presence of hydrogen chloride and hydrogen in the injected gases is also necessary. Since excess hydrogen is added to the carrier gas transporting gallium trichloride from the external boat, hydrogen chloride formed in reaction (\ref{eq:GaCl3-decomposition}) and the remainder of the excess hydrogen are always present in the internal tube of the injector during growth. The calculation of the relative saturation field near the injector nozzle is shown in Fig. \ref{cfd}e. The relative saturation on the surface of the injector $S <1$, which corresponds to the observed suppressed parasitic growth. Growth campaigns with total deposited GaN layer thickness of 5.8~mm were conducted, showing no signs of parasitic deposition on the injector.

During the growth process, the substrate can be rotated at a rate of 10--100 rpm, typical rotation rate used during bulk crystal growth process was 20 rpm. The purpose of the substrate rotation was to average minor inhomogeneities caused by a possible slight displacement of the substrate or substrate holder relative to the axis of symmetry of the reactor or uneven wear of the heating elements. The gas flow pattern remains unaffected by the substrate rotation since, at a rotation rate of 20 rpm, the azimuthal velocity at the edge of the substrate holder is approximately 7~cm/s, which is an order of magnitude lower than the forced gas flow velocity for the typical growth parameters used (Fig. \ref{cfd}b).

\subsection{Growth rate and ammonia partial pressure uniformity}

The advantage of the used injector design is a high achievable homogeneity of the deposition rate and the V-III ratio. The critical factor in achieving high deposition homogeneity in this geometry is the suppression of free convection. The driving force of free convection is the difference in gas density associated with the temperature gradient and the significant difference in the molecular weight of the carrier gas and the precursors.

The predominance of free convection over forced convection is expected at high values of the Rayleigh number~\cite{holstein1992cvd-reactor-review}:
\begin{equation}
Ra=\frac{gL^{3}}{\nu D}\frac{\delta\rho}{\rho} > 1708
\end{equation}
and with a high ratio of the Grashof number to the Reynolds number~\cite{holstein1992cvd-reactor-review}:
\begin{equation}
Gr / Re^2 = \left(\frac{gL^{3}}{\nu^{2}}\frac{\delta\rho}{\rho} \right) / \left( \frac{uL}{\nu}\right)^2\gg 1
\end{equation}
$$ $$
where $L$ is the reactor height, $u$ is the gas flow velocity, $p$ is pressure, $\nu$ is kinematic viscosity, $g$ is gravitational acceleration, and $\delta\rho/\rho$ is the relative inhomogeneity of the gas density associated with inhomogeneity of temperature $\delta T/T$ and gas mixture composition. The nature of the diffusion coefficient $D$  depends on the origin of the density inhomogeneity.  If the density inhomogeneity is determined mainly by the temperature inhomogeneity, then $D$ is the thermal diffusivity, which for nitrogen gas at a temperature of 1000~$^\circ$C is $\approx$2.5~cm$^2$/s.  If the density inhomogeneity is determined mainly by the concentration of a nonuniformly distributed substance with a molecular mass substantially different from the mass of the carrier gas, then $D$ is the diffusion coefficient of this substance. In the described reactor, the temperature gradients in the growth chamber are relatively small and the density inhomogeneity is determined by the concentration of GaCl; the diffusion coefficient of GaCl in nitrogen at T = 1000~$^\circ$C is $\approx 1.3$~cm$^2$/s.

Rayleigh number depends quadratically on pressure:
\begin{equation}
Ra \sim p^2
\end{equation}
and the $Gr/Re^2$ scales quadratically with pressure and the mass flow rate Q:
\begin{equation}
Gr/Re^2 \sim \frac{p^2}{Q^2}
\end{equation}
therefore reducing the process pressure and increasing total gas flow rate are an effective method of suppressing free convection. At the given growth chamber dimensions and the gas flow rate, no free convection is observed at pressures below 200~mbar. 

The gas flow in the growth chamber of the reactor was calculated, taking into account heat transfer, chemical species diffusion, and surface reactions. The total process pressure was 150~mbar; the substrate temperature was 1000~$^\circ$C. The Reynolds number was $Re\approx 30$, Rayleigh number was $Ra\approx 500$ and $Gr/Re^2\approx 0.5$, that corresponded to laminar flow without recirculation and any evidences of free convection (Fig.~\ref{cfd}b). The calculated concentration fields of GaCl and ammonia are shown in Fig.~\ref{cfd}c,d. Ammonia is distributed uniformly over the substrate. A uniform boundary layer depleted with GaCl is observed near the surface of the substrate, where the lines of equal concentration of GaCl are almost parallel to the surface of the substrate. The estimated inhomogeneity of deposition rate and ammonia concentration  along the substrate diameter does not exceed 1\%, as shown in Fig.~\ref{cfd}f.

\begin{figure*}[htb]
	\includegraphics[width=\textwidth]{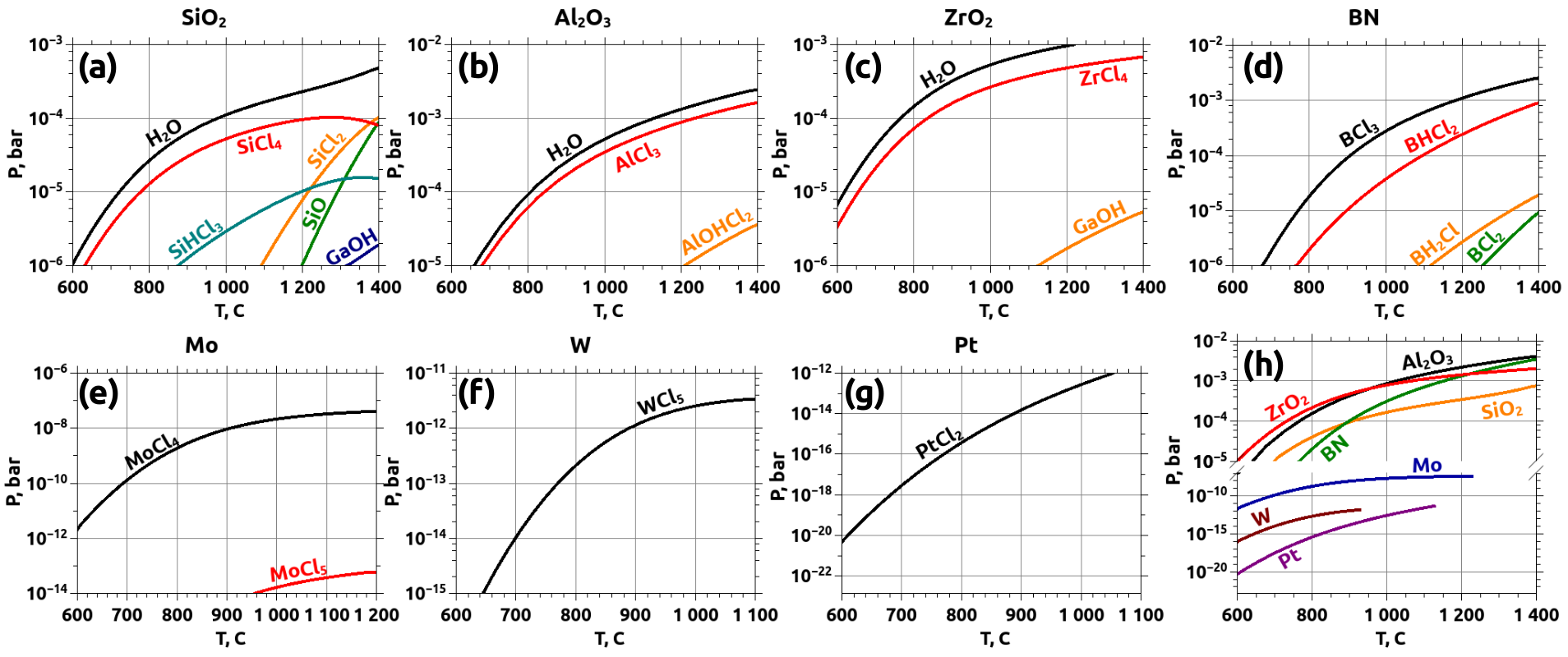}
	\caption{Equilibrium partial pressure of gaseous decomposition products of various materials in a mixture of GaCl:H$_2$:HCl in a ratio of 1:1:1 at atmospheric pressure: (a)~quartz, (b)~alumina, (c)~zirconia, (d)~boron nitride, (e)~molybdenum, (f)~tungsten, (g)~platinum; (h) summary diagram: total partial pressure of gaseous decomposition products over each of the materials. }
	\label{materials}
\end{figure*}

\subsection{Materials}

Special attention was paid to the choice of reactor materials. Quartz, besides being a potential source of contamination, is fragile and limits the ability to produce parts of complex shape and high accuracy. The chemical resistance of commonly used refractory materials was studied using chemical equilibrium analysis. For example, the chemical resistance of quartz, aluminum oxide, zirconium oxide, boron nitride, molybdenum, tungsten and platinum in the gallium chloride injector nozzle was estimated by calculating the equilibrium reaction products in a model mixture containing gallium monochloride, hydrogen, and hydrogen chloride in a ratio of 1:1:1 (Fig. \ref{materials}).

As can be seen, quartz and other oxide materials have equilibrium pressures of decomposition products of more than 10$^{-4}$~atm at a temperature of 1000~$^\circ$C (Fig.~\ref{materials}a,b,c). The main decomposition products are water vapor and metal chloride. Replacing quartz with boron nitride can significantly reduce possible oxygen contamination; however, the equilibrium pressure of BCl$_3$ at 1000~$^\circ$C is still higher than 10$^{-4}$~atm (Fig.~\ref{materials}d). 

Molybdenum, tungsten, and platinum exhibit a significantly lower pressure of the decomposition products in the presence of hydrogen (Fig.~\ref{materials}e,f,g). It should be noted that in the absence of hydrogen, molybdenum and tungsten are not resistant to HCl \cite{brewer1950chemistry}.

Replacing quartz with refractory metals and BN allows excluding one of the potential sources of silicon and oxygen contamination, extend the service life of the injector nozzle, and add flexibility in the design of the growth chamber. 

A detailed description of the thermodynamic study of the chemical resistance of materials, as well as a SIMS study of the concentration of impurities in the metals and ceramics used, precursors, and the resulting GaN layers will be published separately.
\subsection{Exhaust line heating to prevent clogging with ammonium chloride}
\begin{figure}[htb]
	\includegraphics[width=\columnwidth]{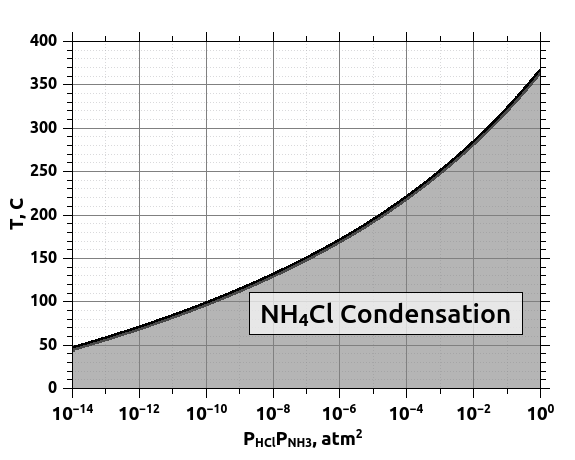}
	\caption{The dependence of the sublimation temperature of ammonium chloride on the partial pressures of ammonia and hydrogen chloride.}
	\label{nh4cl}
\end{figure}
The gases at the exit of the growth chamber contain hydrogen chloride and ammonia, which react to form solid ammonium chloride with decreasing temperature: $$ \rm HCl + NH_3 = NH_4Cl $$ The condensation of ammonium chloride in the exhaust leads to reactor clogging and must be prevented to ensure long continuous growth processes. Heating the exhaust lines down to the powder trap prevents premature condensation. 

The condensation temperature can be defined as
\begin{equation}
T_{NH_{4}Cl}^{cond}=\frac{-\left(\mu_{NH_{4}Cl}^{0}-\mu_{NH_{3}}^{0}-\mu_{HCl}^{0}\right)}{R\ln(P_{HCl}P_{NH_{3}})}.
\end{equation}
where $p_{HCl}$ and $p_{NH_3}$ are the partial pressure of GaCl and ammonia, respectively, and $\mu_{NH_3}$, $\mu_{HCl}$, $\mu_{NH_4Cl}$ are the standard chemical potentials of the species.
The dependence of the temperature required to prevent $\rm NH_4Cl$ condensation on the partial pressures of hydrogen chloride and ammonia is shown in Fig. \ref{nh4cl}.

If gallium chloride does not completely react in the growth chamber, the formation of the GaCl$_3$:NH$_3$ complex is possible in the exhaust. This substance has a boiling point of 438~$^\circ$C \cite{friedman1950observations} and requires significantly higher heating of the exhaust compared to ammonium chloride to prevent condensation.

\section{Load-lock and growth chamber \textit{in-situ} cleaning}
The vacuum load-lock chamber and the procedure of the growth chamber dry etching do not directly affect the crystal growth process, but improve the reproducibility of the process and prolong the service life of the growth chamber parts and the heater, and significantly accelerate and facilitate the workflow. 
The load-lock chamber is located in the lower part of the reactor, where the substrate holder is moved before and after the growth process for loading and unloading. 
Load-lock allows the reactor to be reloaded without exposing the growth chamber to the atmosphere and without thermal cycling.
The use of the load-lock also allowed to implement \textit{in-situ} dry etching of the growth chamber before the growth campaign and between the growth processes. The etching is performed at a temperature of 1000--1100~$^\circ$C in a mixture of hydrogen chloride with hydrogen and allows to remove the GaN deposits and the contaminants that have volatile chlorides.

\section{Bulk GaN layers}
\begin{figure}
	\includegraphics[width=\columnwidth]{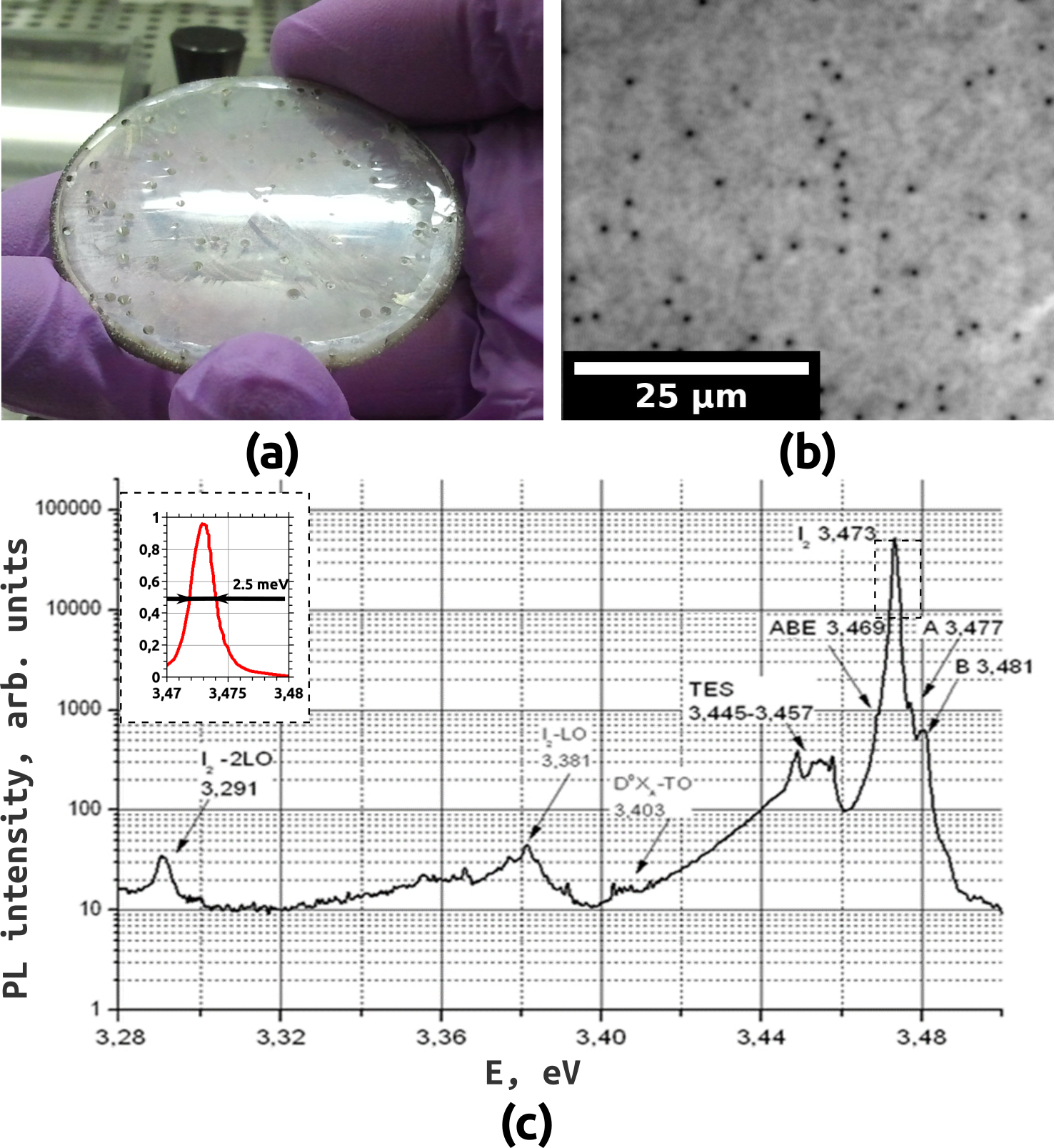}
	\caption{
		(a) Photograph of a freestanding GaN crystal with a thickness of 2800~$\mu$m and a diameter of 50~mm.
		 (b) Cathodoluminescence micrograph of the Ga-face surface. The dark spot density is below $2\cdot10^{6}$ cm$^{-2}$
		 (c) Low temperature (10~K) photoluminescence spectrum of the Ga-face of the crystal.
	 }
	\label{GaN}
\end{figure}
Bulk GaN crystal with a thickness of 2.8~mm, grown on a GaN-on-sapphire template using the two-stage growth process \cite{VVVvoronenkov2013two}, is shown in Fig. \ref{GaN}a. The growth rate was 200 $\mu$m/h, the GaCl:HCl:H$_2$ ratio was 1:2:2.
The crystal separated from the substrate during cooling after growth by spontaneous stress-induced self-separation \cite{williams2007moustakas,voronenkov2019thick}. The as-grown surface was smooth, and the density of V-shaped pits was less than 3~cm$^{-2}$. Microscopic examination showed that the v-pits were associated with macrodefects of the seed template and that no new pits were formed during the bulk layer growth.
The (0002) X-Ray rocking curve FWHM was 51 arcsec. Threading dislocation density estimated using cathodoluminescent microscopy by measuring dark spot density was $2\cdot10^6 \rm{cm}^{-2}$ (Fig.~\ref{GaN}b). The optical quality of the crystal was estimated by photoluminescence (PL). Low temperature (10~K) PL spectrum of the Ga-face of the crystal is shown in Fig.~\ref{GaN}c. The dominant emission line at 3.473~eV that is usually attributed to the exciton bound to a neutral donor \cite{pakula1995growth}, has an FWHM of 2.5~meV.

\section{Conclusions}
The developed reactor has proven itself in long-term operation, with more than a thousand growth processes carried out. Vertical geometry with stagnation point flow provided high uniformity for both the growth rate and the III/V ratio. Vacuum load-lock and \textit{in-situ} growth chamber etching proved to be useful in improving process repeatability. High-temperature chemical-resistant materials of the growth chamber used instead of quartz made it possible to increase the service life and provide an opportunity to reduce the unintentional doping level. High-capacity external precursor sources, gas injector with suppressed parasitic growth, and the heated exhaust line allowed to grow bulk layers with thickness up to 3~mm and higher in a single growth process. 

\section*{Acknowledgment}
The authors gratefully thank Dr.Sci. Valery~Davydov from Ioffe Institue for low-temperature PL measurements, and Prof.~Oleg~Vyvenko and Dr.~Oleg~Medvedev from St.~Petersburg University for CL microscopic studies.

\bibliographystyle{IEEEtran}
\bibliography{hvpe}

\begin{thebibliography}{10}
\providecommand{\url}[1]{#1}
\csname url@samestyle\endcsname
\providecommand{\newblock}{\relax}
\providecommand{\bibinfo}[2]{#2}
\providecommand{\BIBentrySTDinterwordspacing}{\spaceskip=0pt\relax}
\providecommand{\BIBentryALTinterwordstretchfactor}{4}
\providecommand{\BIBentryALTinterwordspacing}{\spaceskip=\fontdimen2\font plus
\BIBentryALTinterwordstretchfactor\fontdimen3\font minus
  \fontdimen4\font\relax}
\providecommand{\BIBforeignlanguage}[2]{{%
\expandafter\ifx\csname l@#1\endcsname\relax
\typeout{** WARNING: IEEEtran.bst: No hyphenation pattern has been}%
\typeout{** loaded for the language `#1'. Using the pattern for}%
\typeout{** the default language instead.}%
\else
\language=\csname l@#1\endcsname
\fi
#2}}
\providecommand{\BIBdecl}{\relax}
\BIBdecl

\bibitem{fujikura2017sciocs}
\BIBentryALTinterwordspacing
H.~Fujikura, T.~Konno, T.~Yoshida, and F.~Horikiri, ``{Hydride-vapor-phase
  epitaxial growth of highly pure {GaN} layers with smooth as-grown surfaces on
  freestanding {GaN} substrates},'' \emph{Japanese Journal of Applied Physics},
  vol.~56, no.~8, p. 085503, jul 2017. [Online]. Available:
  \url{https://doi.org/10.7567%2Fjjap.56.085503}
\BIBentrySTDinterwordspacing

\bibitem{grzegory2006crystallization}
\BIBentryALTinterwordspacing
I.~Grzegory, B.~Lucznik, M.~Bockowski, B.~Pastuszka, G.~Kamler, G.~Nowak,
  M.~Krysko, S.~Krukowski, and S.~Porowski, ``{Crystallization of GaN by HVPE
  on pressure grown seeds},'' \emph{physica status solidi (a)}, vol. 203,
  no.~7, pp. 1654--1657, 2006. [Online]. Available:
  \url{http://doi.org/10.1002/pssa.200565296}
\BIBentrySTDinterwordspacing

\bibitem{bockowski2016challenges}
\BIBentryALTinterwordspacing
M.~Bockowski, M.~Iwinska, M.~Amilusik, M.~Fijalkowski, B.~Lucznik, and
  T.~Sochacki, ``{Challenges and future perspectives in HVPE-GaN growth on
  ammonothermal GaN seeds},'' \emph{Semiconductor Science and Technology},
  vol.~31, no.~9, p. 093002, 2016. [Online]. Available:
  \url{http://doi.org/10.1088/0268-1242/31/9/093002}
\BIBentrySTDinterwordspacing

\bibitem{imanishi2017-hvpe-on-na-flux}
\BIBentryALTinterwordspacing
M.~Imanishi, T.~Yoshida, T.~Kitamura, K.~Murakami, M.~Imade, M.~Yoshimura,
  M.~Shibata, Y.~Tsusaka, J.~Matsui, and Y.~Mori, ``{Homoepitaxial Hydride
  Vapor Phase Epitaxy Growth on GaN Wafers Manufactured by the Na-Flux
  Method},'' \emph{Crystal Growth \& Design}, vol.~17, no.~7, pp. 3806--3811,
  2017. [Online]. Available: \url{https://doi.org/10.1021/acs.cgd.7b00388}
\BIBentrySTDinterwordspacing

\bibitem{Tietjen01071966}
\BIBentryALTinterwordspacing
J.~J. Tietjen and J.~A. Amick, ``{The Preparation and Properties of
  Vapor-Deposited Epitaxial GaAs$_{1-x}$P$_x$ Using Arsine and Phosphine},''
  \emph{Journal of The Electrochemical Society}, vol. 113, no.~7, pp. 724--728,
  1966. [Online]. Available:
  \url{http://jes.ecsdl.org/content/113/7/724.abstract}
\BIBentrySTDinterwordspacing

\bibitem{maruska1969}
\BIBentryALTinterwordspacing
H.~P. Maruska and J.~J. Tietjen, ``{The preparation and properties of
  vapor-deposited single-crystalline GaN},'' \emph{Applied Physics Letters},
  vol.~15, no.~10, pp. 327--329, 1969. [Online]. Available:
  \url{https://doi.org/10.1063/1.1652845}
\BIBentrySTDinterwordspacing

\bibitem{safvi1997kuech}
\BIBentryALTinterwordspacing
S.~Safvi, N.~Perkins, M.~Horton, R.~Matyi, and T.~Kuech, ``{Effect of reactor
  geometry and growth parameters on the uniformity and material properties of
  GaN/sapphire grown by hydride vapor-phase epitaxy},'' \emph{Journal of
  Crystal Growth}, vol. 182, no.~3, pp. 233 -- 240, 1997. [Online]. Available:
  \url{http://www.sciencedirect.com/science/article/pii/S0022024897003758}
\BIBentrySTDinterwordspacing

\bibitem{tavernier2000clarke}
\BIBentryALTinterwordspacing
P.~R. Tavernier, E.~V. Etzkorn, Y.~Wang, and D.~R. Clarke, ``{Two-step growth
  of high-quality GaN by hydride vapor-phase epitaxy},'' \emph{Applied Physics
  Letters}, vol.~77, no.~12, pp. 1804--1806, 2000. [Online]. Available:
  \url{https://aip.scitation.org/doi/abs/10.1063/1.1311600}
\BIBentrySTDinterwordspacing

\bibitem{dam2004hageman-horizontal}
\BIBentryALTinterwordspacing
C.~Dam, A.~Grzegorczyk, P.~Hageman, R.~Dorsman, C.~Kleijn, and P.~Larsen,
  ``{The effect of HVPE reactor geometry on GaN growth rate-experiments versus
  simulations},'' \emph{Journal of Crystal Growth}, vol. 271, no.~1, pp. 192 --
  199, 2004. [Online]. Available:
  \url{http://www.sciencedirect.com/science/article/pii/S0022024804009042}
\BIBentrySTDinterwordspacing

\bibitem{segal2004illegems}
\BIBentryALTinterwordspacing
A.~Segal, A.~Kondratyev, S.~Karpov, D.~Martin, V.~Wagner, and M.~Ilegems,
  ``{Surface chemistry and transport effects in GaN hydride vapor phase
  epitaxy},'' \emph{Journal of Crystal Growth}, vol. 270, no.~3, pp. 384 --
  395, 2004. [Online]. Available:
  \url{http://www.sciencedirect.com/science/article/pii/S0022024804008401}
\BIBentrySTDinterwordspacing

\bibitem{monemar2005}
\BIBentryALTinterwordspacing
B.~Monemar, H.~Larsson, C.~Hemmingsson, I.~Ivanov, and D.~Gogova, ``{Growth of
  thick GaN layers with hydride vapour phase epitaxy},'' \emph{Journal of
  Crystal Growth}, vol. 281, no.~1, pp. 17 -- 31, 2005, the Internbational
  Workshop on Bulk Nitride Semiconductors III. [Online]. Available:
  \url{http://www.sciencedirect.com/science/article/pii/S002202480500299X}
\BIBentrySTDinterwordspacing

\bibitem{richter2006aitron-horizontal}
\BIBentryALTinterwordspacing
{E. Richter and Ch. Hennig and M. Weyers and F. Habel and J.-D. Tsay and W.-Y.
  Liu and P. Brückner and F. Scholz and Yu. Makarov and A. Segal and J.
  Kaeppeler}, ``{Reactor and growth process optimization for growth of thick
  GaN layers on sapphire substrates by HVPE},'' \emph{Journal of Crystal
  Growth}, vol. 277, no.~1, pp. 6 -- 12, 2005. [Online]. Available:
  \url{http://www.sciencedirect.com/science/article/pii/S0022024804021232}
\BIBentrySTDinterwordspacing

\bibitem{williams2007moustakas}
\BIBentryALTinterwordspacing
A.~D. Williams and T.~Moustakas, ``{Formation of large-area freestanding
  gallium nitride substrates by natural stress-induced separation of GaN and
  sapphire},'' \emph{Journal of Crystal Growth}, vol. 300, no.~1, pp. 37 -- 41,
  2007, first International Symposium on Growth of Nitrides. [Online].
  Available:
  \url{http://www.sciencedirect.com/science/article/pii/S0022024806011857}
\BIBentrySTDinterwordspacing

\bibitem{fujito2009mchemical}
\BIBentryALTinterwordspacing
K.~Fujito, S.~Kubo, H.~Nagaoka, T.~Mochizuki, H.~Namita, and S.~Nagao, ``{Bulk
  GaN crystals grown by HVPE},'' \emph{Journal of Crystal Growth}, vol. 311,
  no.~10, pp. 3011 -- 3014, 2009, proceedings of the 2nd International
  Symposium on Growth of III Nitrides. [Online]. Available:
  \url{http://www.sciencedirect.com/science/article/pii/S002202480900089X}
\BIBentrySTDinterwordspacing

\bibitem{amilusik2013topgan}
\BIBentryALTinterwordspacing
{M. Amilusik and T. Sochacki and B. Łucznik and M. Boćkowski and B. Sadovyi
  and A. Presz and I. Dziecielewski and I. Grzegory}, ``{Analysis of
  self-lift-off process during HVPE growth of GaN on MOCVD-GaN/sapphire
  substrates with photolitographically patterned Ti mask},'' \emph{Journal of
  Crystal Growth}, vol. 380, pp. 99 -- 105, 2013. [Online]. Available:
  \url{http://www.sciencedirect.com/science/article/pii/S0022024813004028}
\BIBentrySTDinterwordspacing

\bibitem{fleischmann2019influence}
\BIBentryALTinterwordspacing
S.~Fleischmann, E.~Richter, A.~Mogilatenko, M.~Weyers, and G.~Tr{\"a}nkle,
  ``{Influence of quartz on silicon incorporation in HVPE grown AlN},''
  \emph{Journal of Crystal Growth}, vol. 507, pp. 295--298, 2019. [Online].
  Available: \url{https://doi.org/10.1016/j.jcrysgro.2018.11.028}
\BIBentrySTDinterwordspacing

\bibitem{hemmingsson2008modeling}
\BIBentryALTinterwordspacing
C.~Hemmingsson, G.~Pozina, M.~Heuken, B.~Schineller, and B.~Monemar,
  ``{Modeling, optimization, and growth of GaN in a vertical halide vapor-phase
  epitaxy bulk reactor},'' \emph{Journal of Crystal Growth}, vol. 310, no.~5,
  pp. 906--910, 2008. [Online]. Available:
  \url{https://doi.org/10.1016/j.jcrysgro.2007.11.062}
\BIBentrySTDinterwordspacing

\bibitem{villars1959method}
\BIBentryALTinterwordspacing
D.~S. Villars, ``{A method of successive approximations for computing
  combustion equilibria on a high speed digital computer},'' \emph{The Journal
  of Physical Chemistry}, vol.~63, no.~4, pp. 521--525, 1959. [Online].
  Available: \url{http://doi.org/10.1021/j150574a016}
\BIBentrySTDinterwordspacing

\bibitem{colonna2004hierarchical}
\BIBentryALTinterwordspacing
G.~Colonna and A.~D'angola, ``{A hierarchical approach for fast and accurate
  equilibrium calculation},'' \emph{Computer Physics Communications}, vol. 163,
  no.~3, pp. 177--190, nov 2004. [Online]. Available:
  \url{http://doi.org/10.1016/j.cpc.2004.08.004}
\BIBentrySTDinterwordspacing

\bibitem{colonna2007improvements}
\BIBentryALTinterwordspacing
G.~Colonna, ``{Improvements of hierarchical algorithm for equilibrium
  calculation},'' \emph{Computer Physics Communications}, vol. 177, no.~6, pp.
  493--499, 2007. [Online]. Available:
  \url{http://doi.org/10.1016/j.cpc.2007.01.012}
\BIBentrySTDinterwordspacing

\bibitem{vvv-thesis}
\BIBentryALTinterwordspacing
V.~Voronenkov, ``{Epitaxial growth of bulk gallium nitride layers:
  technological parameters optimization (in russian).}'' Ph.D. dissertation,
  Ioffe Institute, 2015. [Online]. Available:
  \url{http://doi.org/10.12731/AAAA-B15-415121030009-4}
\BIBentrySTDinterwordspacing

\bibitem{gurvich1990thermodynamic}
L.~V. Gurvich and I.~Veyts, \emph{{Thermodynamic Properties of Individual
  Substances: Elements and Compounds}}.\hskip 1em plus 0.5em minus 0.4em\relax
  CRC press, 1990.

\bibitem{nist-janaf1998}
M.~W. Chase, Ed., \emph{{NIST-JANAF Themochemical Tables, Fourth
  Edition}}.\hskip 1em plus 0.5em minus 0.4em\relax American Chemical Society,
  1998.

\bibitem{przhevalskii1998thermodynamic}
\BIBentryALTinterwordspacing
I.~Przhevalskii, S.~Y. Karpov, and Y.~N. Makarov, ``{Thermodynamic properties
  of group-III nitrides and related species},'' \emph{MRS Internet Journal of
  Nitride Semiconductor Research}, vol.~3, pp. 1--16, 1998. [Online].
  Available: \url{http://doi.org/10.1557/s1092578300001022}
\BIBentrySTDinterwordspacing

\bibitem{zinkevich2004thermodynamic}
\BIBentryALTinterwordspacing
M.~Zinkevich and F.~Aldinger, ``{Thermodynamic Assessment of the Gallium-Oxygen
  System},'' \emph{Journal of the American Ceramic Society}, vol.~87, no.~4,
  pp. 683--691, 2004. [Online]. Available:
  \url{http://doi.org/10.1111/j.1551-2916.2004.00683.x}
\BIBentrySTDinterwordspacing

\bibitem{schafer1980mocl345}
H.~Schafer, ``{Studien zum chemischen Transport von MoCl$_3$ Ein Beitrag zur
  chemischen Thermodynamik der Molybdnchloride},'' \emph{Zeitschrift fur
  anorganische und allgemeine Chemie}, vol. 469, no.~1, pp. 123--127, 1980.

\bibitem{schonherr1988ptcl-growth}
\BIBentryALTinterwordspacing
E.~Schonherr, M.~Wojnowski, A.~Rabenau, and S.~Lacher, ``{On the growth of
  PtCl$_3$, crystals from the vapour phase},'' \emph{Journal of the Less-Common
  Metals}, vol. 137, pp. 211--286, 1988. [Online]. Available:
  \url{https://doi.org/10.1016/0022-5088(88)90093-8}
\BIBentrySTDinterwordspacing

\bibitem{brewer1950chemistry}
L.~Brewer, L.~A. Bromley, P.~W. Gilles, and N.~L. Lofgren, \emph{{Chemistry and
  Metallurgy of miscellaneous materials: Thermodynamics}}.\hskip 1em plus 0.5em
  minus 0.4em\relax New York, McGraw-Hill, 1950.

\bibitem{popinet2003gerris}
\BIBentryALTinterwordspacing
S.~Popinet, ``{Gerris: a tree-based adaptive solver for the incompressible
  Euler equations in complex geometries},'' \emph{Journal of Computational
  Physics}, vol. 190, no.~2, pp. 572--600, 2003. [Online]. Available:
  \url{https://doi.org/10.1016/S0021-9991(03)00298-5}
\BIBentrySTDinterwordspacing

\bibitem{seifert1981study}
\BIBentryALTinterwordspacing
W.~Seifert, G.~Fitzl, and E.~Butter, ``{Study on the growth rate in VPE of
  GaN},'' \emph{Journal of Crystal Growth}, vol.~52, pp. 257--262, 1981.
  [Online]. Available: \url{https://doi.org/10.1016/0022-0248(81)90201-3}
\BIBentrySTDinterwordspacing

\bibitem{malinovsky1982growth}
\BIBentryALTinterwordspacing
V.~V. Malinovsky, L.~A. Marasina, I.~G. Pichugin, and M.~Tlaczala, ``{The
  Growth Kinetics and Surface Morphology of GaN Epitaxial Layers on
  Sapphire},'' \emph{Crystal Research and Technology}, vol.~17, no.~7, pp.
  835--840, 1982. [Online]. Available:
  \url{http://doi.org/10.1002/crat.2170170708}
\BIBentrySTDinterwordspacing

\bibitem{usui1997bulk}
\BIBentryALTinterwordspacing
A.~Usui, ``{Bulk GaN Crystal With Low Defect Density Grown By Hydride Vapor
  Phase Epitaxy},'' \emph{MRS Proceedings}, vol. {482}, p. 233, 1997. [Online].
  Available: \url{https://doi.org/10.1557/PROC-482-233}
\BIBentrySTDinterwordspacing

\bibitem{fornari2001hydride}
\BIBentryALTinterwordspacing
R.~Fornari, M.~Bosi, N.~Armani, G.~Attolini, C.~Ferrari, C.~Pelosi, and
  G.~Salviati, ``{Hydride vapour phase epitaxy growth and characterisation of
  GaN layers},'' \emph{Materials Science and Engineering: B}, vol.~79, no.~2,
  pp. 159--164, 2001. [Online]. Available:
  \url{http://doi.org/10.1016/s0921-5107(00)00584-5}
\BIBentrySTDinterwordspacing

\bibitem{koukitu2002surface}
\BIBentryALTinterwordspacing
A.~Koukitu, M.~Mayumi, and Y.~Kumagai, ``{Surface polarity dependence of
  decomposition and growth of GaN studied using in situ gravimetric
  monitoring},'' \emph{Journal of Crystal Growth}, vol. 246, no.~3, pp.
  230--236, 2002. [Online]. Available:
  \url{http://doi.org/10.1016/s0022-0248(02)01746-3}
\BIBentrySTDinterwordspacing

\bibitem{ban1972mass_gan}
\BIBentryALTinterwordspacing
V.~S. Ban, ``{Mass Spectrometric Studies of Vapor-Phase Crystal Growth II.}''
  \emph{Journal of the Electrochemical Society}, vol. 119, no.~6, pp. 761--765,
  1972. [Online]. Available: \url{http://doi.org/10.1149/1.2404322}
\BIBentrySTDinterwordspacing

\bibitem{liu1978growth}
\BIBentryALTinterwordspacing
S.~S. Liu and D.~A. Stevenson, ``{Growth kinetics and catalytic effects in the
  vapor phase epitaxy of gallium nitride},'' \emph{Journal of The
  Electrochemical Society}, vol. 125, no.~7, pp. 1161--1169, 1978. [Online].
  Available: \url{https://doi.org/10.1149/1.2131641}
\BIBentrySTDinterwordspacing

\bibitem{miura1995new}
\BIBentryALTinterwordspacing
Y.~Miura, N.~Takahashi, A.~Koukitu, and H.~Seki, ``{New Epitaxial Growth Method
  of Cubic GaN on (100) GaAs Using (CH$_3$)$_3$Ga, HCl and NH$_3$},''
  \emph{Japanese journal of applied physics}, vol.~34, no.~4A, p. L401, 1995.
  [Online]. Available: \url{https://doi.org/10.1143/JJAP.34.L401}
\BIBentrySTDinterwordspacing

\bibitem{kryliouk1999large}
\BIBentryALTinterwordspacing
O.~Kryliouk, M.~Reed, T.~Dann, T.~Anderson, and B.~Chai, ``{Large area GaN
  substrates},'' \emph{Materials Science and Engineering: B}, vol.~66, no. 1-3,
  pp. 26--29, 1999. [Online]. Available:
  \url{https://doi.org/10.1016/S0921-5107(99)00114-2}
\BIBentrySTDinterwordspacing

\bibitem{fahle2013hcl}
\BIBentryALTinterwordspacing
D.~Fahle, D.~Brien, M.~Dauelsberg, G.~Strauch, H.~Kalisch, M.~Heuken, and
  A.~Vescan, ``{HCl-assisted growth of GaN and AlN},'' \emph{Journal of Crystal
  Growth}, vol. 370, pp. 30--35, 2013. [Online]. Available:
  \url{https://doi.org/10.1016/j.jcrysgro.2012.10.027}
\BIBentrySTDinterwordspacing

\bibitem{Lindeke1970-TEG-Cl}
\BIBentryALTinterwordspacing
K.~Lindeke, W.~Sack, and J.~J. Nickl, ``{Gallium Diethyl Chloride: A New
  Substance in the Preparation of Epitaxial Gallium Arsenide},'' \emph{Journal
  of The Electrochemical Society}, vol. 117, no.~10, pp. 1316--1318, 1970.
  [Online]. Available: \url{http://jes.ecsdl.org/content/117/10/1316.abstract}
\BIBentrySTDinterwordspacing

\bibitem{nickl1974preparation}
\BIBentryALTinterwordspacing
J.~Nickl, W.~Just, and R.~Bertinger, ``{Preparation of epitaxial
  galliumnitride},'' \emph{Materials Research Bulletin}, vol.~9, no.~10, pp.
  1413--1420, 1974. [Online]. Available:
  \url{https://doi.org/10.1016/0025-5408(74)90066-X}
\BIBentrySTDinterwordspacing

\bibitem{ZHANG2000-TEG-Cl-GaN}
\BIBentryALTinterwordspacing
L.~Zhang, S.~Gu, T.~Kuech, and M.~P. Boleslawski, ``{Gallium nitride growth
  using diethyl gallium chloride as an alternative gallium source},''
  \emph{Journal of Crystal Growth}, vol. 213, no.~1, pp. 1 -- 9, 2000.
  [Online]. Available:
  \url{http://www.sciencedirect.com/science/article/pii/S0022024800003341}
\BIBentrySTDinterwordspacing

\bibitem{park2005identification}
\BIBentryALTinterwordspacing
C.~Park, J.-h. Kim, D.~Yoon, S.~Han, C.~Doh, S.~Yeo, K.-H. Lee, and T.~J.
  Anderson, ``{Identification of a gallium-containing carbon deposit produced
  by decomposition of trimethyl gallium},'' \emph{Journal of The
  Electrochemical Society}, vol. 152, no.~5, pp. C298--C303, 2005. [Online].
  Available: \url{https://doi.org/10.1149/1.1873452}
\BIBentrySTDinterwordspacing

\bibitem{lee1996vapor}
\BIBentryALTinterwordspacing
H.~Lee and J.~S. Harris~Jr, ``{Vapor phase epitaxy of GaN using GaCl$_3$/N$_2$
  and NH$_3$/N$_2$},'' \emph{Journal of crystal growth}, vol. 169, no.~4, pp.
  689--696, 1996. [Online]. Available:
  \url{https://doi.org/10.1016/S0022-0248(96)00472-1}
\BIBentrySTDinterwordspacing

\bibitem{varadarajan2004chloride}
\BIBentryALTinterwordspacing
E.~Varadarajan, P.~Puviarasu, J.~Kumar, and R.~Dhanasekaran, ``{On the chloride
  vapor-phase epitaxy growth of GaN and its characterization},'' \emph{Journal
  of crystal growth}, vol. 260, no. 1-2, pp. 43--49, 2004. [Online]. Available:
  \url{http://doi.org/10.1016/j.jcrysgro.2003.08.021}
\BIBentrySTDinterwordspacing

\bibitem{yamane2011tri}
\BIBentryALTinterwordspacing
T.~Yamane, K.~Hanaoka, H.~Murakami, Y.~Kumagai, and A.~Koukitu, ``{Tri-halide
  vapor phase epitaxy of GaN using GaCl$_3$ gas as a group III precursor},''
  \emph{physica status solidi c}, vol.~8, no.~5, pp. 1471--1474, 2011.
  [Online]. Available: \url{http://doi.org/10.1002/pssc.201000902}
\BIBentrySTDinterwordspacing

\bibitem{ueda2011effects}
\BIBentryALTinterwordspacing
T.~Ueda, M.~Yuri, and J.~S. Harris~Jr, ``{Effects of Growth Temperatures on
  Crystal Quality of GaN by Vapor Phase Epitaxy Using GaCl$_3$ and NH$_3$},''
  \emph{Japanese Journal of Applied Physics}, vol.~50, no.~8R, p. 085501, 2011.
  [Online]. Available: \url{http://doi.org/10.7567/jjap.50.085501}
\BIBentrySTDinterwordspacing

\bibitem{murakami2016tri}
\BIBentryALTinterwordspacing
H.~Murakami, N.~Takekawa, A.~Shiono, Q.~T. Thieu, R.~Togashi, Y.~Kumagai,
  K.~Matsumoto, and A.~Koukitu, ``{Tri-halide vapor phase epitaxy of thick GaN
  using gaseous GaCl$_3$ precursor},'' \emph{Journal of Crystal Growth}, vol.
  456, pp. 140--144, 2016. [Online]. Available:
  \url{http://doi.org/10.1016/j.jcrysgro.2016.08.029}
\BIBentrySTDinterwordspacing

\bibitem{kobayashi1991low}
\BIBentryALTinterwordspacing
R.~Kobayashi, Y.~Jin, F.~Hasegawa, A.~Koukitu, and H.~Seki, ``{Low temperature
  growth of GaAs and AlAs by direct reaction between GaCl$_3$, AlCl$_3$ and
  AsH$_3$},'' \emph{Journal of crystal growth}, vol. 113, no. 3-4, pp.
  491--498, 1991. [Online]. Available:
  \url{http://doi.org/10.1016/0022-0248(91)90084-i}
\BIBentrySTDinterwordspacing

\bibitem{holstein1992cvd-reactor-review}
\BIBentryALTinterwordspacing
W.~L. Holstein, ``{Design and modeling of chemical vapor deposition
  reactors},'' \emph{Progress in Crystal Growth and Characterization of
  Materials}, vol.~24, no.~2, pp. 111 -- 211, 1992. [Online]. Available:
  \url{http://doi.org/10.1016/0960-8974(92)90031-k}
\BIBentrySTDinterwordspacing

\bibitem{friedman1950observations}
H.~L. Friedman and H.~Taube, ``{Observations on the Chlorogallates and Related
  Compounds},'' \emph{Journal of the American Chemical Society}, vol.~72,
  no.~5, pp. 2236--2243, 1950.

\bibitem{VVVvoronenkov2013two}
\BIBentryALTinterwordspacing
V.~Voronenkov, N.~Bochkareva, R.~Gorbunov, P.~Latyshev, Y.~Lelikov, Y.~Rebane,
  A.~Tsyuk, A.~Zubrilov, U.~Popp, M.~Strafela, and Y.~Shreter, ``{Two modes of
  HVPE growth of GaN and related macrodefects},'' \emph{Physica Status Solidi
  (c)}, vol.~10, no.~3, pp. 468--471, 2013. [Online]. Available:
  \url{http://doi.org/10.1002/pssc.201200701}
\BIBentrySTDinterwordspacing

\bibitem{voronenkov2019thick}
\BIBentryALTinterwordspacing
V.~V. Voronenkov, Y.~S. Lelikov, A.~S. Zubrilov, A.~A. Leonidov, and Y.~G.
  Shreter, ``{Thick GaN film stress-induced self-separation},'' in \emph{2019
  IEEE Conference of Russian Young Researchers in Electrical and Electronic
  Engineering (EIConRus)}.\hskip 1em plus 0.5em minus 0.4em\relax IEEE, 2019,
  pp. 833--837. [Online]. Available:
  \url{http://doi.org/10.1109/eiconrus.2019.8657271}
\BIBentrySTDinterwordspacing

\bibitem{pakula1995growth}
\BIBentryALTinterwordspacing
K.~Pakula, J.~Baranowski, R.~Stepniewski, A.~Wysmolek, I.~Grzegory, J.~Jun,
  S.~Porowski, M.~Sawicki, and K.~Starowieyski, ``{Growth of GaN Metalorganic
  Chemical Vapour Deposition Layers on GaN Single Crystals},'' \emph{Acta
  Physica Polonica-Series A General Physics}, vol.~88, no.~5, pp. 861--864,
  1995. [Online]. Available: \url{http://doi.org/10.12693/aphyspola.88.861}
\BIBentrySTDinterwordspacing

\end{thebibliography}


\end{document}